\documentclass[graybox]{svmult}

\usepackage{mathptmx}       
\usepackage{helvet}         
\usepackage{courier}        
\usepackage{type1cm}        
\usepackage{makeidx}         
\usepackage{graphicx}        
\usepackage{multicol}        
\usepackage[bottom]{footmisc}

\usepackage{pdfsync}
\usepackage{amsfonts,amssymb,amsmath}

\newcommand{\bes}{\begin{displaymath}}
\newcommand{\ees}{\end{displaymath}}
\newcommand{\be}{\begin{equation}}
\newcommand{\ee}{\end{equation}}
\newcommand{\ba}{\begin{eqnarray}}
\newcommand{\ea}{\end{eqnarray}}
\newcommand{\bas}{\begin{eqnarray*}}
\newcommand{\eas}{\end{eqnarray*}}

\newcommand{\B}{{\@Bbb B}}
\newcommand{\C}{{\@Bbb C}}
\newcommand{\T}{{\mathbb T}}
\newcommand{\F}{{\@Bbb F}}

\newcommand{\Q}{{\@Bbb Q}}
\newcommand{\bQ}{{\@Bbb Q}}
\newcommand{\N}{{\@Bbb N}}
\newcommand{\R}{{\mathbb R}}

\newcommand{\bbR}{{\mathbb R}}
\newcommand{\W}{{\@Bbb W}}
\newcommand{\Z}{{\mathbb Z}}
\newcommand{\bbZ}{{\Z}}

\newcommand{\al}{\alpha}

\newcommand{\eps}{\epsilon}

\newcommand{\cA}{\@s A}
\newcommand{\cB}{\@s B}
\newcommand{\cC}{\@s C}
\newcommand{\cD}{\@s D}
\newcommand{\cE}{\@s E}
\newcommand{\cF}{\@s F}
\newcommand{\cG}{\@s G}
\newcommand{\cH}{\@s H}
\newcommand{\cI}{\@s I}
\newcommand{\cJ}{\@s J}
\newcommand{\mc}{\mathcal}
\newcommand{\cK}{\@s K}
\newcommand{\cL}{\@s L}
\newcommand{\cN}{\@s N}
\newcommand{\cM}{\@s M}
\newcommand{\cO}{\@s O}
\newcommand{\cP}{\@s P}
\newcommand{\cR}{\@s R}
\newcommand{\cS}{\@s S}
\newcommand{\cT}{\@s T}
\newcommand{\cV}{\@s V}
\newcommand{\cW}{\@s W}
\newcommand{\cX}{\@s X}
\newcommand{\cY}{\@s Y}
\newcommand{\cZ}{\@s Z}

\newcommand{\bma}{\@bm a}
\newcommand{\bmb}{\@bm b}
\newcommand{\bmc}{\@bm c}
\newcommand{\bmd}{\@bm d}

\newcommand{\bme}{\@bm e}
\newcommand{\bmf}{\@bm f}
\newcommand{\bmg}{\@bm g}
\newcommand{\bmh}{\@bm h}
\newcommand{\bmi}{\@bm i}
\newcommand{\bmj}{\@bm j}
\newcommand{\bmk}{\@bm k}
\newcommand{\bml}{\@bm l}
\newcommand{\bmm}{\@bm m}
\newcommand{\bmn}{\@bm n}
\newcommand{\bmo}{\@bm o}
\newcommand{\bmp}{\@bm p}
\newcommand{\bmq}{\@bm q}
\newcommand{\bmr}{\@bm r}
\newcommand{\bms}{\@bm s}
\newcommand{\bmt}{\@bm t}
\newcommand{\bmu}{\@bm u}
\newcommand{\bmw}{\@bm w}
\newcommand{\bmv}{\@bm v}
\newcommand{\bmx}{\@bm x}
\newcommand{\bx}{\@bm x}

\newcommand{\bmy}{\@bm y}
\newcommand{\bz}{\@bm z}

\newcommand{\bU}{{\bf U}}

\newcommand{\by}{\@bm y}

\newcommand{\bmzero}{\@bm 0}

\newcommand{\gA}{\@g A}
\newcommand{\gD}{\@g D}
\newcommand{\gJ}{\@g J}
\newcommand{\gF}{\@g F}
\newcommand{\gM}{\@g M}
\newcommand{\gR}{\@g R}
\newcommand{\gu}{\@g u}
\newcommand{\gr}{\@g r}
\newcommand{\gp}{\@g p}

\makeindex             

\begin{document}

\title*{Thermal conductivity in harmonic lattices with random collisions.}
\titlerunning{Conductivity of harmonic lattices} 
\author{Giada Basile, C\'edric Bernardin, Milton Jara, Tomasz Komorowski and Stefano Olla}
\authorrunning{Basile, Bernardin, Jara, Komorowski, Olla} 
\institute{Giada Basile \at  Dipartimento di Matematica,
 Universit\`{a} di Roma La Sapienza,
 Roma, Italy\\
  \email{basile@mat.uniroma.it} 
\and C\'edric Bernardin \at Laboratoire J.A. Dieudonn\'e UMR CNRS 7351,
Universit\'e de Nice Sophia-Antipolis, Parc Valrose, 
06108 Nice Cedex 02, France\\
\email{cbernard@unice.fr} 
\and Milton Jara \at IMPA, Rio de Janeiro, Brazil\\
\email{mjara@impa.br}
\and Tomasz Komorowski \at Institute of Mathematics, Polish Academy of
Sciences, Warsaw, Poland\\
\email{komorow@hektor.umcs.lublin.pl}
\and Stefano Olla \at Ceremade, UMR CNRS 7534,  Universit\'e Paris
Dauphine, 75775 Paris Cedex 16, France\\
\email{olla@ceremade.dauphine.fr} 
}

%
%
\maketitle
\date{30-11-2015  
}
\abstract*{We review recent rigorous mathematical results about the
  macroscopic behaviour of harmonic chains with the dynamics perturbed
  by a random exchange of velocities between nearest neighbor
  particles. The random exchange models the effects of nonlinearities
of anharmonic chains and the resulting dynamics have similar
macroscopic behaviour. In particular there is a superdiffusion of
energy for unpinned acoustic chains. The corresponding evolution of the
temperature profile is governed by a fractional heat equation. In
non-acoustic chains we have normal diffusivity, even if momentum is
conserved.} 

\abstract{We review recent rigorous mathematical results about the
  macroscopic behaviour of harmonic chains with the dynamics perturbed
  by a random exchange of velocities between nearest neighbor
  particles. The random exchange models the effects of nonlinearities
of anharmonic chains and the resulting dynamics have similar
macroscopic behaviour. In particular there is a superdiffusion of
energy for unpinned acoustic chains. The corresponding evolution of the
temperature profile is governed by a fractional heat equation. In
non-acoustic chains we have normal diffusivity, even if momentum is
conserved. } 

\section{Introduction}
\label{sec:introduction}

Lattice systems of coupled anharmonic oscillators have been widely used
in order to understand the macroscopic transport of the energy,
in particular the superdiffusive behavior in one and two dimensional
unpinned chains. While a lot of numerical experiments and {heuristic}
considerations have been made (cf. \cite{llp97, sll, Sp13} and many
contributions in the present volume), very few mathematical rigourous
scaling limits have been obtained until now. 

For harmonic chains it is possible to perform explicit computations,
even in the stationary state driven by thermal boundaries
(cf. \cite{rll}). But since these dynamics are completely integrable,
the energy transport is purely ballistic and they do not provide help
in understanding the 
diffusive or superdiffusive behavior of anharmonic chains.

The \emph{scattering} effect of the non-linearities can be modeled by
stochastic perturbations of the dynamics such that they conserve total
momentum and total energy, like a random exchange of the velocities
between {the nearest neighbor particles}. 
We will describe the results for the one-dimensional chains, and we
will mention the results in {the higher dimensions} in Section
\ref{sec:model-d-dimensions}. In particular we will prove how the
transport through the fractional Laplacian, either asymmetric or
symmetric, emerges from microscopic models. 

The \emph{infinite dynamics} is described by the velocities  and
positions 
{$\{(p_x, q_x)\in\mathbb R^2\}_{x\in\mathbb Z}$} of the particles.
The formal Hamiltonian is given by  
\begin{equation}
\label{ham}
{\cal H}({p},{q}):=\frac{1}{2m}\sum_{x}{p}_x^2 + 
\frac{1}{2}\sum_{x,x'}\alpha_{x-x'}{q}_x{q}_{x'},
\end{equation}
where we assume {that} the masses are equal to $1$ 
and that $\alpha$ is symmetric with a finite range or at most
exponential decay $|\alpha_x|\le C e^{-c|x|}$.
 We define the Fourier transform of a function $f:\Z\to \R$ as $ \hat
 f(k) = \sum_x e^{-2\pi i x k} f(x)$ for $k\in \T$. We assume that
 $\hat\alpha(k) >0$ for $k\neq0$. The function $\omega(k) =
 \sqrt{\hat\alpha(k)}$ is called the {\em dispersion 
relation} of the chain.

\paragraph{Unpinned Chains} 

We are particularly interested in the
unpinned chain, i.e. $\hat\alpha(0) = 0$, when {the} total momentum is
conserved even under the stochastic dynamics described below.
 {Then,} the infinite system is translation 
 invariant under shift in q, and the correct coordinates are the
 interparticle distances (also called {stretches, or strains}):
 \begin{equation}
   \label{eq:2}
   r_x = q_x - q_{x-1}, \qquad x\in\Z.
 \end{equation}
When $\hat\alpha''(0)>0$ we say that the chain is \emph{acoustic}
(i.e. there is a non-vanishing sound speed). We
will see that this is a crucial condition for the superdiffusivity of
the energy in one dimension. 
For  unpinned acoustic chains we have that $\omega(k) \sim |k|$ as
$k\to 0$.

\paragraph{Pinned chains}
When $\hat\alpha(0) > 0$, the system is pinned and translation
invariance is broken. In this case $\omega(k) \sim k^2$ as $k\to 0$. 
This is also the case for unpinned non-acoustic chains, and this is
responsable for the diffusive behavior of the energy, cf. Section 
\ref{sec:non-acoustic-chains}. 

\paragraph{Dynamics with Stochastic Collisions}
{To the Hamiltonian} dynamics we add random elastic collisions, where
momenta of the nearest--neighbor particles are exchanged. This happens at
independent random exponential times: each couple of particles
{labeled} ${x,x+1}$ exchange their velocities $p_x$ and $p_{x+1}$ 
at exponential independent random times of intensity $\gamma$. 
Equivalently there are independent Poisson processes $\{N_{x,x+1}(t),\;
x\in \mathbb Z\}$ of intensity
$\gamma$, independent from the positions and velocities of all
particles.    
The evolution of the system is described by the stochastic
differential equations
\begin{equation}
  \label{eq:sdeN}
  \begin{split}
    \dot q_x &= p_x \\
    \dot p_x &= -(\alpha * q(t))_x + \left(p_{x+1}(t^-) -
      p_x(t^-)\right) \dot N_{x,x+1}(t) + \left(p_{x-1}(t^-) -
      p_x(t^-)\right) \dot N_{x-1,x}(t)
  \end{split}
\end{equation}
where $\dot N_{x,x+1}(t) = \sum_j \delta(t- T_{x,x+1}(j))$,
with $\{T_{x,x+1}(j)\}$ the random times when $N_{x,x+1}$ jumps, and
$p_x(t^-)$ is the 
velocity of the particle $x$ just before time $t$, i.e.
$\lim_{s\downarrow 0} p_x(t-s)$. 

The evolution of the probability density on the configurations then
follows the Fokker-Planck equations:
\begin{equation}
  \label{eq:FP}
  \partial_t f(t,p,q) = \left(A+S\right) f(t,p,q)
\end{equation}
where $A$ is the hamiltonian operator
\begin{equation}
  \label{eq:A}
  A = \sum_x \left(p_x \partial_{q_x} - (\partial_{q_x} \mathcal H) 
    \partial_{p_x} \right),
\end{equation}
while $S$ is the generator of the random exchanges
\begin{equation}
  \label{eq:S}
  S f (p,q) = \gamma\sum_x \left(f(p^{x,x+1},q) - f(p,q) \right),
\end{equation}
where $p^{x,x+1}$ is the configuration obtained exchanging $p_x$ and $p_{x+1}$.

This stochastic perturbation of the Hamiltonian dynamics has the
property to conserve the total energy, and in the unpinned case the
resulting dynamics {conserves} also the total momentum
($\sum_x p_x$), and the 
\emph{volume} or \emph{strain} of the chain ($\sum_x r_x$). It has
also the property that these are the only conserved quantities. In this
sense it gives the necessary ergodicity to the dynamics  (\cite{FFL, bo14}). 

We have also considered other type of conservative random dynamics,
like a continuous random exchange of the momenta of
each triplets $\{p_{x-1}, p_x, p_{x+1}\}$. The intersection of the
kinetic energy sphere $p_{x-1}^2 + p_x^2 + p_{x+1}^2=C$ with the plane
$p_{x-1}+ p_x + p_{x+1}=C'$ gives a one dimensional circle. Then we
define a dynamics on this circle by a standard Wiener process on the
corresponding angle. This perturbation is locally more mixing, but   
it gives the same results for the macroscopic transport.

\paragraph{Equilibrium stationary measures}

Due to the harmonicity of the interactions, the Gibbs equilibrium
stationary measures are Gaussians. Positions and momenta are
independent and the distribution is parametrized, accordingto teh
rules of statistical mechanics, by the temperature
$T=\beta^{-1}>0$.   
In the pinned case, they are formally given by
$$
\nu_{\beta} (dq, dp) \sim \frac{e^{- \beta {\mc H} (p,q)}}{\mathcal
  Z}\;  dq\, dp.
$$
In the unpinned case, the correct definition {should involve} the
$r_x$ variables.  Then the distribution of the $r_x$'s is Gaussian and
becomes uncorrelated in the case of the nearest neighbor interaction.  
For \emph{acoustic unpinned chains} the Gibbs measures are parameterized by 
$$
\boldsymbol\lambda=
(\beta^{-1} \text{(temperature)}, \bar p \text{(velocity)}, \tau
\text{(tension)}),
$$ 
and are given formally by 
\begin{equation}
  \label{eq:gibbs1}
  \nu_{\boldsymbol\lambda} (dr, dp) \sim 
\frac{
e^{- \beta\,  [ {\mc H} (p,q) -{\bar p} \sum_x p_x -\tau \sum_x r_x ]}}
{\mathcal Z}  \; dr\, dp.
\end{equation}
Non-acoustic chains are tensionless and the equilibrium measures have a
different parametrization, see Section \ref{sec:non-acoustic-chains}.

\paragraph{Macroscopic space-time scales}
We will mostly concentrate on the acoustic unpinned case (except
Section \ref{sec:non-acoustic-chains}).
In this case the total hamiltonian can be written as
\begin{equation}
\label{ham2}
{\cal H}({p},{q}):=\sum_{x} \frac{p_x^2}2 - 
\frac{1}{4}\sum_{x,x'}\alpha_{x-x'}\left({q}_x-{q}_{x'}\right)^2,
\end{equation}
where $q_x - q_{x'} = \sum_{y=x'+1}^{x} r_y$ for $x>x'$.
We define the energy per atom as:
\begin{equation}
  \label{eq:energy}
  e_x ({r}, {p}):= \frac{p_x^2}2 - \frac 14 \sum_{x'}
  \alpha_{x-x'} (q_x -  q_{x'})^2.
\end{equation}

There are three conserved {(also called
\emph{balanced}) fields}: the energy $\sum_x e_x$, the momentum
$\sum_x p_x$ and the strain $\sum_x r_x$ 
. We want to study
the macroscopic evolution of the spatial distribution of these
fields in a large space-time scale. We introduce a scale parameter
$\eps>0$ and, for any smooth test function $J:\R \to \R$, define the
empirical distribution 
\begin{equation}
  \label{eq:4}
  \eps \sum_x J(\eps x) \bU_x(\eps^{-a} t) , \qquad \bU_x =
  (r_x, p_x, e_x). 
\end{equation}

  We are interested in the limit as $\eps\to 0$. {The} parameter
  $a \in [1,2]$ corresponds to different possible scalings. The
  value $a = 1$ corresponds to the hyperbolic {scaling}, while $a = 2$
  corresponds to the diffusive scaling. 
  The intermediate values $1< a<  2$ {pertain to} the superdiffusive scales.  

The interest of the unpinned model is that there are three different
macroscopic space-scales where we {observe} non-trivial behaviour
{of the chain}: $a = 1, \tfrac{3}{2} , 2$.   

\section{Hyperbolic scaling: the linear wave equation.}
\label{sec:hyperb-scal-line}

Let us assume that the dynamics starts with a random initial
distribution $\mu_\eps=\langle \cdot \rangle_\eps$ of finite energy of size $\eps^{-1}$, i.e. for
some positive constant $E_0$:
\begin{equation}
  \label{eq:1}
  \eps \sum_{x} \langle e_x(0) \rangle_\eps \ \le\ E_0.
\end{equation}
where we denote $e_x(t) = e_x(p(t),q(t))$.
We will also assume that some smooth macroscopic initial profiles
  $$
\frak  u_0(y) = (\frak r_0(y), \frak p_0(y), \frak e_0(y))
$$ 
are associated  with  the initial distribution, in the sense that: 
\begin{equation}
  \label{eq:3}
  \begin{split}
\lim_{\eps\to 0} \mu_\eps\left( \left|\eps \sum_x J(\eps x) \bU_x(0)  -
  \int J(y) \frak u_0(y) dy\right| > \delta \right) = 0, \qquad
\forall \delta>0,
  \end{split}
\end{equation}
for any test function $J$. 

Then it can be proven {(\cite{koyper})} that
these initial profiles are governed by the linear wave equation in
the following sense 
\begin{equation}
  \label{eq:5}
  \lim_{\eps\to 0} \eps \sum_x J(\eps x) \bU_x(\eps^{-1} t)  = \int J(y) 
   \frak u(y,t) dy
\end{equation}
where ${\frak  u(y,t)} =(\frak r(y,t),\frak p(y,t),\frak e(y,t))$  is the
solution of: 
\begin{equation}
  \label{eq:linw}
  \begin{split}
    \partial_t \frak r = \partial_y \frak p, \qquad 
    \partial_t \frak p = \tau_1\partial_y \frak r, \qquad 
    \partial_t \frak e = \tau_1 \partial_y (\frak p \frak r)
  \end{split}
\end{equation}
with $\tau_1= \frac{\hat\alpha''(0)}{8\pi^2}$ (the square of the speed
of sound of the chain).  Notice that in the non-acoustic case 
$\hat\alpha''(0) = 0$, there is no evolution at the hyperbolic scale.

Observe that the evolution of the fields of strain {${\frak r}$} and momentum
{${\frak p}$} is autonomous of the energy
field. Furthermore we can define the 
macroscopic \emph{mechanical} energy as
\begin{equation}
  \label{eq:8}
  \frak e_{mech}(y,t) = \frac 12 \left( \tau_1 \frak r(y,t)^2 + \frak
    p(y,t)^2\right) 
\end{equation}
and the temperature profile or \emph{thermal} energy as
\begin{equation}
  \label{eq:7}
  T(y,t) =\frak e(y,t) - \frak e_{mech}(y,t).
\end{equation}
It follows immediately from \eqref{eq:linw} that $T(y,t) = T(y,0)$,
i.e. the temperature profile does not change on the hyperbolic
space-time scale. 

There is a corresponding decomposition of the energy of the random
initial configurations: long wavelengths (invisible for the exchange
noise in the dynamics) contribute to the mechanical
energy and they will evolve in this hyperbolic scale following the
linear equations \eqref{eq:linw}. This energy will eventually disperse
to infinity at large time (in this scale). Because of the noise
dynamics, short wavelength will contribute to the variance
(temperature) of the distribution, and the correponding profile does
not evolve in this hyperbolic scale. See \cite{koyper} for the details
of this decomposition.

\section{Superdiffusive evolution of the temperature profile.}
\label{sec:superd-evol-temp}

As we have seen in the previous section, for acoustic chains the
mechanical part of the 
energy evolves ballistically in the hyperbolic scale and eventually it
will disperse to infinity. Consequently when we look at the larger
superdiffusive time scale $\eps^{-a} t$, $a >1$, we start
{only with the} thermal profile of energy, while the strain and momentum
profiles are equal to $0$. It turns out that, for acoustic chains, the
temperature profile evolves at the time scale correspondig to $a=3/2$:
\begin{equation}
  \label{eq:super1}
  \lim_{\eps\to 0} \eps \sum_x J(\eps x)\;  \langle e_x(\eps^{-3/2} t) \rangle_{\eps}  = 
  \int J(y) \frak T(y,t) dy, \qquad t>0, 
\end{equation}
where $\frak T(y,t)$ solves the fractional heat equation
\begin{equation}
  \label{eq:fhe}
  \partial_t \frak T = -c|\Delta_y|^{3/4} \frak T, \qquad \frak T(y,0)
  =  T(y,0)
\end{equation}
where $c = \alpha''(0)^{3/4} 2^{-9/4}(3\gamma)^{-1/2}$. 
This is proven in \cite{jko2}.
We also have that the profiles for the other conserved quantities
remain flat: 
  \begin{equation}
  \label{eq:super2}
  \begin{split}
    \lim_{\eps\to 0} \eps \sum_x J(\eps x) \; \langle r_x(\eps^{-3/2}
    t) \rangle_{\eps} 
    &= 0,\\
  \lim_{\eps\to 0} \eps \sum_x J(\eps x) \; \langle p_x(\eps^{-3/2} t)
  \rangle_{\eps}  &= 0, \quad t>0. 
  \end{split}
\end{equation}
Clearly the null value is due to the finite energy assumption,
otherwise it will be the corresponding constants, i.e.
$\lim_{\eps\to 0}  \eps \sum_x \langle r_x(0) \rangle_{\eps}$ and  
$\lim_{\eps\to 0}  \eps \sum_x \langle p_x(0) \rangle_{\eps}$. 


In finite volume with given boundary conditions (periodic or else),
the mechanical energy will persist, oscillating in linear waves. At
the larger superdiffusive time scale, waves will oscillate fast giving
a 
weak convergence for the initial profiles of strain and momentum to
constant values.

\section{The diffusive behavior of the phonon-modes.}
\label{sec:diff-behav-phon}

We have seen that \eqref{eq:super2} holds at the superdiffusive time scale
and consequently at any larger time scale. But if we recenter the
evolution of the strain and momentum around the propagation of the
wave equation we see {Gaussian fluctuations} at the diffusive space time
scale. For this purpose it is useful to introduce a microscopic
approximation of the Riemann invariants (normal modes) of the wave
equation:
\begin{equation}
 f^{\pm}_x(t)\ =\ p_x(t)\ \pm\ \tau_1^{1/2} \left( r_x(t)\
        {\pm \ \frac {3\gamma -1}2 (r_{x+1}(t) - r_x(t))}
      \right).
\label{eq:9}
    \end{equation}

Once $f^{\pm}_x(t)$ are recentred on the Riemann invariants of
the wave equation, they diffuse on the proper space-time scale, more
precisely:  
\begin{equation*}
        \begin{split}
          \eps \sum_x J(\eps x \mp \tau_1^{1/2} \eps^{-1} t) &\; f^{\pm}_x(\eps^{-2} t)\
        \mathop{\longrightarrow}_{\eps\to 0} \ \int_\R J(y)\;
       {\bar{\frak f}^{\pm,d}(y,t) dy},\\
        \partial_t\; \bar{\frak f}^{\pm,d} &= \
        \frac{3\gamma}2 \ \partial^2_y \;\bar{\frak f}^{\pm,d}.
      \end{split}
      \end{equation*}
For a proof see \cite{koyper}.

\section{Equilibrium fluctuations}
\label{sec:equil-fluct}

If we start with an equilibrium stationary measure corresponding to a
certain temperature $T$, momentum $\bar p$ and strain $\bar r$, then
of course there will be no evolution of the empirical fields defined
by \eqref{eq:4}. 
But {we shall observe} the evolution of the equilibrium time correlations
defined by 
\begin{equation}
  \label{eq:etc}
  S^{\ell,\ell'}(x,t) = \left<u^\ell_x(t) \, u^{\ell'}_0(0) \right> -
  \left<u^\ell_0(0)\right> \left< u^{\ell'}_0(0) \right> , \qquad u^1_x =
  r_x,\ u^2_x=p_x,\ u^3_x = e_x,
\end{equation}
where $\left<\cdot\right>$ denotes the expectation with respect to the
dynamics in the corresponding equilibrium.  Let us assume for
simplicity of notation that $\bar r = 0 =\bar p$, otherwise we have to
shift $x$ along the characteristics of the linear wave equation. 
At time $t=0$, it is
easy to compute the limit (in a distributional sense):
\begin{equation}
  \label{eq:13}
  \lim_{\eps\to 0} \eps^{-1/2} S([\eps^{-1} y],0) = {\delta(y)}
  \begin{pmatrix}
    T & 0 & \tilde\alpha T\\
    0 & T & 0\\
    \tilde\alpha T & 0 & (\frac 12 + \alpha^2)T^2 
  \end{pmatrix}
\end{equation}
where $\tilde\alpha$ is a constant depending only on the interaction.

In the hyperbolic time scale $a =1$ this correlation matrix evolves
deterministically, i.e.
\begin{equation}
  \label{eq:12}
   \lim_{\eps\to 0} \eps^{-1/2} S([\eps^{-1} y],\eps^{-1} t) = \bar S(y,t)
\end{equation}
where 
\begin{equation}
  \label{eq:10}
  \partial_t \bar S^{11}(y,t) = \partial_y  \bar S^{22}(y,t), \qquad 
  \partial_t  \bar S^{22}(y,t) = \tau_1\partial_y  \bar S^{11}(y,t),
\end{equation}
while for the energy correlations
\begin{equation}
  \label{eq:6}
  \partial_t \bar  S^{33}(y,t) = 0. 
\end{equation}
In particular if $\bar p = \bar r = 0$, energy fluctuations do not
evolve at the hyperbolic time scale. As for the energy profile out of
equilibrium, the evolution is at a further time scale. 
By a duality argument (cf. \cite{jko2}), the
evolution of the energy correlations  {occurs} at the
superdiffusive time scale {with $a = 3/2$, namely} 
\begin{equation}
  \label{eq:122}
   \lim_{\eps\to 0} \eps^{-1/2} S^{33}([\eps^{-1} y],\eps^{-3/2} t) =
   \tilde S^{33}(y,t) ,
\end{equation}
{where $\tilde S^{33}(y,t) $}
is the solution of:
\begin{equation}
  \label{eq:11}
   \partial_t  \tilde S^{33} = - c |\Delta_y|^{3/4}\tilde S^{33}.
\end{equation}

\section{The phonon Boltzmann equation}

A way to understand the energy superdiffusion in the one-dimensional
system is to analyze its kinetic limit, i.e. a limit for weak noise
where the number of stochastic collisions per unit time remains
bounded. In the non-linear case it corresponds to a weak non-linearity
limit as proposed first 
in his seminal paper \cite{Pe}
in 1929 by Peierls. 
He intended  to compute thermal conductivity for insulators in analogy
with the kinetic theory of gases. 
The main idea is that at low temperatures the lattice vibrations
responsible {for the} energy transport can be
described as  a gas of interacting particles (phonons) characterized
by a wave number $k$. The time-dependent distribution function of phonons
 solves a Boltzmann type equation. Over the last years, starting from the work of Spohn \cite{Sp06}, several papers are devoted to achieve   
phononic Boltzmann type equations from microscopic dynamics of
oscillators.
 
A rigorous derivation can be achieved for the chain of harmonic
oscillators perturbed by a 
stochastic exchange of velocities \cite{BOS}. 
The main tool is the introduction of a Wigner function, which
describes the energy density of the phonons. 
Let $\hat\psi$ be the complex field
$$
\hat\psi(k)=\frac 1 {\sqrt 2}\big(\omega(k)\hat q (k)+
i\hat p(k)\big),\qquad k\in\mathbb T, 
$$
where $\hat p$, $\hat q$ are the Fourier transform of the variables
$p$, $q$ and $\omega(k)$ is the dispersion relation of chain.  
The energy of the chain can be expressed in terms of the fields
$\hat\psi$, namely 
$
{\mathcal H}=\int_{\mathbb T} dk\,| \hat\psi(k)|^2.
$
The evolution of the field $\hat\psi$ due to the  the pure harmonic
Hamiltonian without noise ($\gamma = 0$) reads
$$
\partial_t\hat\psi(k,t)=-i\omega(k)\hat\psi(k,t),
$$
therefore the quantities $|\hat\psi(k)|^2$ are preserved by the
harmonic dynamics.  

The Wigner distribution is defined in analogy to the usual one in
quantum mechanics  
\begin{equation}
  \begin{split}
   & W^\varepsilon(y,k,t)=(\varepsilon/2)\int_{\mathbb R} e^{{2\pi }i\xi y}
    \widehat{W}^{\varepsilon}(\xi,k,t) d\xi ,\label{eq:wigner}\\
   & \widehat{W}^{\varepsilon}(\xi,k,t):= \langle\hat\psi(k-\varepsilon
    \xi/2,t)^*\; \hat\psi(k+\varepsilon\xi/2,t) \rangle_\varepsilon
  \end{split}
\end{equation}
where $\langle \cdot\rangle_\varepsilon$ denotes the expectation value
with respect to the initial measure,  chosen in such a way that   the
average of  the total energy,
 is of order $\varepsilon^{-1}$, i.e.  
{$\varepsilon \,\langle {\mathcal H}\rangle_\varepsilon \le E_0$, see
  \eqref{eq:1}}. We also require that 
the averages of all $p_x$ and $q_x$ are zero. 

The Wigner distribution $ W^\varepsilon(y,k,t)$ defined by
\eqref{eq:wigner} {gives} a different energy distribution 
{from the one considered} in the previous sections,
i.e. $<e_{[\varepsilon^{-1}y]}(t)>$. 
But in the macroscopic limit, as $\varepsilon \to 0$, they are
equivalent \cite{jko2}. 


We look {at} the evolution of the Wigner function on a time scale
$\varepsilon^{-1}t$, with the strength of the noise  of order
$\varepsilon$, i.e. we consider the dynamics defined by
\eqref{eq:sdeN} with $\gamma$ replaced by $\varepsilon \gamma$, like
in a
Boltzmann-Grad limit.   
This evolution is not autonomous and is given by
\begin{equation}\label{BEeps}\begin{split}
\partial_t \hat{W}^{\varepsilon}(\xi,k,t)=&
 -i\xi\, \omega'(k) \hat{W}^{\varepsilon}(\xi,k,t)
 +\gamma \, \mathcal C \hat{W}^{\varepsilon}(\xi,k,t)\\
 &-\frac \gamma 2 \, \mathcal C \hat{Y}^{\varepsilon}(\xi,k,t)
-\frac \gamma 2 \, \mathcal C \hat{Y}^{\varepsilon}(\xi,k,t)^* 
+\mathcal O(\varepsilon),
\end{split}\end{equation}
where $\hat{Y}^{\varepsilon}$ is the field
$\hat{Y}^{\varepsilon}(\xi,k,t)=\langle\hat\psi(k+\varepsilon
\xi/2,t)\hat\psi(k-\varepsilon\xi/2,t) \rangle_\varepsilon$ and $\mathcal C$ is
a linear operator. The transport term is due to the harmonic
hamiltonian, while the "collision"  operator $\mathcal C$ is related to the
stochastic noise.

It turns out that the field $\hat{Y}^{\varepsilon}$ is the Wigner distribution
associated to the \emph{difference} between the kinetic energy and the
potential energy. This is fast oscillating on the time scale
$\varepsilon^{-1}$  and in the limit $\varepsilon\to 0$ it disappears
after time integration. 
Therefore  in the limit $\varepsilon\to 0$ the Wigner function
$W^\varepsilon$ weakly converges to the solution of the following
linear Boltzmann equation 
\begin{equation}
\label{BE}
\partial_t W(y,k,t)+ {\frac 1 {2\pi}}
\omega'(k)\, \partial_y W(y,k,t)= \mathcal C W(y,k,t).
\end{equation}
It describes the evolution of the energy density distribution, over
the physical space $\mathbb R$,  of the   phonons, characterized by a
wave number $k$ and traveling with velocity $\omega'(k)$. 
We remark that for the unpinned acoustic chains $\omega'(k)$ remains
strictly positive for small $k$
. 
The collision term has the following expression
$$
\mathcal C f(k)=\int_{\mathbb T} dk' R(k,k')[f(k')-f(k)],
$$
where the kernel $R$ is positive and symmetric.
{One can write an exact expression on $R$, nevertheless its} crucial
feature is that   $R$ behaves like $ k^2$ for small $k$,  due to the
fact that the noise preserves the total momentum.  
Na\"ively, it means  that 
phonons with small wave numbers 
travel with a finite velocity, but they have low probability to be
scattered, thus   their mean free paths 
have a macroscopic length (ballistic transport). This intuitive
picture has an exact statement in the probabilistic interpretation of
\eqref{BE}. 
The equation describes the evolution of the
probability density of a  Markov process $\left(K(t),
Y(t)\right)$ on $\mathbb T\times\mathbb R$, where
$K(t)$ is a reversible jump process  and $Y(t)$ is an
additive functional of $K$, namely $Y(t)=\int_0^t \omega'(K_s)ds.$
A phonon with {the} wave number $k$ waits in its state
an exponentially distributed random time $\tau(k)$ with mean value 
$\sim k^{-2}$ for small $k$. Then it jumps to another state $k'$ with
probability $\sim k'^2 dk'$.  The additive functional $Y(t)$ describes
the position of the phonon and can be expressed as
$$
Y(t)=\sum_{i=1}^{\mathcal N_t}\tau(X_i)\omega'(X_i),
$$
where $\{  X_i\}_{i\geq 1}$ is the Markov chain given by the sequence
of the states visited by the process $K(t)$. Here $\mathcal N_t$
denotes the number of jumps up to the time $t$. 
The  tail distribution of the random variables
$\{\tau(X_i)\omega'(X_i)\}$  with respect to the stationary measure
{$\pi$} of the chain behaves like 
$$
\pi\left[\vert  \tau(X_i)\omega'(X_i)\vert>\lambda\right]\sim
\frac{1}{\lambda^{3/2}}.
$$
Therefore the variables $\tau(X_i)\omega'(X_i)$ have an infinite variance
with respect to the stationary measure.  We remark that the variance
 is exactly the expression of the thermal conductivity obtained in
 \cite{bbo2}. The rescaled process
$N^{-2/3}Y(N {t})$ converges in distribution to a stable symmetric L\'evy
process with index $3/2$ (\cite{jko}, \cite{babo}). As a
corollary, the rescaled solution of the Boltzmann equation
$W(N^{2/3}y, k, Nt)$ converges, {as $N\to+\infty$}, to the solution of
the fractional diffusion equation 
$$
\partial_t u(y,t)=-\big|\Delta|^{3/4} u(y,t).
$$
A different, more analytic approach can be found in \cite{mmm}.

In the pinned or the non-acoustic cases,  $\omega'\sim k$ for small $k$,
$\tau(X_i)\omega'(X_i)$ has finite variance 
with respect to the stationary measure. In particular this variance
 coincides with the thermal diffusivity computed by the Green-Kubo formula
\cite{bbo2}. Then one can prove that the rescaled solution
$W(N^{1/2}y, k, Nt)$ converges to the heat equation 
$$
\partial_t u(y,t)= D \Delta u(y,t),
$$ 
 with $D$ given by the thermal diffusivity.

The results described above give a two step approach to the
diffusion or superdiffusion of the energy: first a
{kinetic} limit where the Boltzmann phonon equation is obtained in the
weak noise limit, 
then a  {superdiffusive} or diffusive rescaling of the
solution of this equation. 
The  {results} described in Section \pageref{sec:superd-evol-temp} concern
a simple space-time rescaling, without any weak noise
approximation\footnote{In the non-linear cases we cannot expect that
  the two step approach would give the same result of the direct
  rescaling of the dynamics. 
  In the $\beta$-FPU the kinetic limit seems to
give a different superdiffusion scaling than the direct limit
\cite{Luk-spohn,Sp13}.}.   
Still Boltzmann equation helps to understand the proof
of \eqref{eq:super1}, that goes under the following lines.

Let us just consider the superdiffusive case and consider the evolution
of the Wigner distribution at time $\varepsilon^{-3/2}t$ and for {the} noise
of {intensity} $\gamma$.  {Equation \eqref{BEeps} becomes}
\begin{equation}
  \label{eq:BEeps2}
  \begin{split}
\partial_t \hat{W}^{\varepsilon}(\xi,k,t)=&
 -i\xi\, \omega'(k) \varepsilon^{-1/2} \hat{W}^{\varepsilon}(\xi,k,t)
 +\gamma  \varepsilon^{-3/2} \, \mathcal C\hat{W}^{\varepsilon}(\xi,k,t)\\
 &-\frac \gamma 2 \varepsilon^{-3/2} \, 
 \mathcal C\left(\hat{Y}^{\varepsilon}(\xi,k,t) + 
 \hat{Y}^{\varepsilon}(\xi,k,t)^*\right) +\mathcal O(\varepsilon),
\end{split}
\end{equation}
This looks like a very singular limit. Still {due to fast oscillations }
 the $\hat{Y}^{\varepsilon}$ terms disappears after
time integration. Furthermore,  {because the number}
of collisions 
per unit time tends to infinity, the limit of the Wigner distribution
\emph{homogenizes} in the variable $k$, i.e. its limit {becomes} a function
$W(y,t)$ {independent of variable} $k$. 
Assume {that the above facts have been proven}, and consider the case
of the simple random exchange of the velocities, that give a
scattering rate of the form: 
$R(k,k')= R(k)R(k')$, with $\int R(k) dk =1$. The argument below, that
follows the line of \cite{mmm}, can be
generalized to {various} rate functions \cite{jko2}.
{The Laplace transform $w_\varepsilon$ in time of the Wigner distribution satisfies}
the equation: 
\begin{equation}
  \label{eq:24}
  \begin{split}
    \left(\lambda + 2\gamma\varepsilon^{-3/2} R(k) + i\omega'(k) \xi
      \varepsilon^{-1/2}\right) w_\varepsilon(\lambda, \xi, k) \\
    = \widehat W_\eps(\xi,k,0) + \varepsilon^{-3/2} 2\gamma R(k) \int
    R(k') dk' + \mathcal O ({\varepsilon}).
  \end{split}
\end{equation}
By dividing the above expression by $D_\varepsilon(\lambda,\xi,k) =
\varepsilon^{3/2}\lambda + 2\gamma R(k) + i\omega'(k) \xi
\varepsilon$, multiplying by $2\gamma R(k)$ and  
 integrating in $k$ one obtains
\begin{equation}
  \label{eq:14}
  a_\varepsilon \int  w_\varepsilon(\lambda, \xi, k) R(k) dk - \int
  \frac{2\gamma R(k)\widehat
    W_\eps(\xi,k,0)}{D_\varepsilon(\lambda,\xi,k)} dk\ =\  \mathcal O({\varepsilon})
\end{equation}
where, since $R(k) \sim k^2$ and {due to} the
assumptions  {made} on $\omega'(k)$: 
\begin{equation*}
  a_\varepsilon\ =\ 2\gamma \epsilon^{-3/2} \left( 1 - \int \frac{2\gamma
    R^2(k)}{ D_\varepsilon(\lambda,\xi,k)} dk\right)
   \ \mathop{\longrightarrow}_{\varepsilon\to 0}\ \lambda + c |\xi|^{3/2}.
\end{equation*}
{Thanks to} the homogenization {property} of $\widehat W_\varepsilon$
\begin{equation*}
  \int  w_\varepsilon(\lambda, \xi, k) R(k) dk
  \ \mathop{\longrightarrow}_{\varepsilon\to 0}\ w(\lambda,\xi).
\end{equation*}
Furthermore 
$$
\frac{2\gamma R(k)}{D_\varepsilon(\lambda,\xi,k)}\
 \mathop{\longrightarrow}_{\varepsilon\to 0}\ 1,
$$
and we conclude that the limit function {$w(\lambda,\xi)$} satisfies
the equation 
\begin{equation}
  \label{eq:18}
  \left(\lambda + c |\xi|^{3/2}\right) w(\lambda,\xi)\ =\ \int
  \widehat W(\xi,k,0) dk
\end{equation}
where $\widehat W(\xi,k,0)$ is the limit of the initial condition. 
Equation \eqref{eq:18} is the Laplace-Fourier transform of the
fractional heat equation. Assuming instead that $\omega'(k) \sim k$, a
similar argument gives the normal heat equation \cite{jko2}.

\section{Non-acoustic chains: beam dynamics}
\label{sec:non-acoustic-chains}

We have seen in the previous sections that the ballistic behaviour at
the hyperbolic scale and the superdiffusive behavior of the energy 
strictly  depends on the positivity of the sound velocity $\tau_1$. 
In the case {$\hat\alpha(0) =\hat\alpha''(0) =0$} all the coefficients
in these evolutions are null and in fact the {limit of the energy
  follows a regular diffusion}. Notice 
that the dynamics is still momentum conserving. 

A typical example is given by the following choice of the interaction:
\begin{equation}
  \label{eq:15}
{\alpha_0 = 3/2, \quad \alpha_1 = -1, \quad \alpha_2= 1/4,\quad
  \alpha_x = 0,\quad x>2}  
\end{equation}
{(then $\hat\al(k)=4\sin^4(\pi k)$) that corresponds} to the Hamiltonian
\begin{equation}
  \label{eq:16}
  \mathcal H = \sum_x \left[ \frac {p_x^2}2 + 
{\frac12}(q_{x+1} - q_x)^2
      - {\frac{ 1}{8}} (q_{x+2} -  q_x)^2\right] =  
   \sum_x \left[ \frac {p_x^2}2 +  {\frac{ 1}{8}}
     \left(q_{x+1} - 2q_x +  q_{x-1}\right)^2\right]. 
\end{equation}
Notice that the expression \eqref{eq:gibbs1} is not defined for any
value of the tension $\tau$, this is why we also call these chains
{tensionless}. Basically, the equilibrium energy does not change by
\emph{pulling} the chain. It does change by \emph{bending} it, this is
why the relevant quantities are defined by  
the local \emph{curvatures} or \emph{deflections}:
\begin{equation}
  \label{eq:curva}
  {\kappa}_x = -\Delta q_x = 2q_x - q_{x+1} - q_{x-1}. 
\end{equation}
The relevant balanced quantities are now $({\kappa}_x, p_x, e_x)$. 

The invariant equilibrium measures are formally given by  
\begin{equation}
  \label{eq:nagibbs}
  \frac{e^{-\beta \left[\mathcal H - \bar p \sum_x p_x - \mathfrak L \sum_x
          {\kappa}_x \right]}}{\mathcal Z} \, d\kappa \, dp
\end{equation}
where the parameter $\mathfrak L$ is called \emph{load}. Notice that
these measures would be non-translation invariant {in the} coordinates
$r_x$'s. 

Under these conditions the sound velocity $\tau_1$ is always null, and
there is no ballistic evolution of the chain. In fact it turns out that the
macroscopic evolution of the three quantities 
{$(\frak k(t,y), \frak p(t,y), \frak e(t,y))$}
is diffusive. By defining 
$$
\mathfrak e_{mech}(t,y) = \frac 12 \left( \frak p(t,y)^2 + \frac 14
  \frak k^2\right) 
\quad  \text{and}\quad
\mathfrak T(t,y) = \mathfrak e(t,y) - \mathfrak e_{mech}(t,y),
$$
after the corresponding space-time scaling {we obtain
(\cite{kononac}): 
\begin{equation}
  \label{eq:17}
  \left\{
    \begin{array}{l}
    {  \partial_t \frak k\ =-\ \partial_y^2 \frak p}, \\
      \partial_t \frak p\ =\
      {\frac{1}{4}} \partial_y^2 \frak k + 
      {\gamma} \partial_y^2 \frak p,\\
      \partial_t \mathfrak T \ =\ 
      D_\gamma \partial_y^2 \mathfrak T
      +\frac{\gamma}{2}\left[\left(\partial_y{\frak 
      p}\right)^2-{\frak p}\partial_y^2{\frak p}\right].
          \end{array}
\right.
\end{equation}}
The first two equations are the damped Euler-Bernoulli beam
equations. The third {one} describes the diffusive behavior of the
energy. The thermal diffusivity $D_\gamma$ can be computed explicitely
and diverges as $\gamma\to 0$ (the deterministic dynamics has
ballistic energy transport as every harmonic chain).  
(the damping  terms {involving $\gamma$} are due to the exchange noise). In
particular for constant initial values of $\frak k$ and  $\frak p$,
the energy (temperature) profile follows a normal heat equation with {the}
thermal diffusivity that can be computed explicitly
(cf. \cite{kononac}). 

These models provide {rigorous  counter-examples} to the usual
conjecture that {the} momentum conservation in one dimension always  implies
superdiffusivity of the energy (cf \cite{sll,dhar}). The presence of a
non-vanishing sound velocity seems a necessary condition.


\section{A simpler model with two conserved quantities}
\label{sec:simpler-model-with}

In this section we consider the nearest neighbors unpinned harmonic
chain (with mass $1$ and coupling forces
$\alpha_1=\alpha_{-1}=-\tfrac{\alpha_0}{2}$) but we add different
stochastic collisions with the properties that they conserve the total
energy and some extra quantity (that we call "volume") but no longer
momentum and stretch. By defining 
$a =\sqrt{\alpha_0}$
and the field 
$\{\eta_x \in \R\; ;\; x\in \Z\}$ by $\eta_{2x} = a r_x$ and
$\eta_{2x+1} = p_x$, the Hamiltonian equations are reduced to  
$$\dot {\eta_x} = a (\eta_{x+1} -\eta_{x-1}).$$ 
The stochastic collisions are such that at random times given by
independent Poisson clocks $N_{x,x+1} (t)$ of intensity $\gamma$ the
kinetic energy at site $x$ is exchanged with the corresponding
potential energy. The simplest way to do it is to exchange the
variable $\eta_x$ with $\eta_{x+1}$. Because of the form of the noise
the total energy $\sum_x \tfrac{\eta_x^2}{2m}$ and the "volume"
$\sum_x \eta_x = \sum_x (p_x + ar_x)$ are the only conserved
quantities of the dynamics (\cite{BS}). By 
reducing the number of conserved quantities from $3$ (energy,
momentum, stretch) to $2$ (energy, volume) we expect to see easily the
influence of the other conserved quantity on the superdiffusion of
energy. The nature of the superdiffusion for models with two conserved
quantities are studied in the nonlinear fluctuating hydrodynamics
framework by Spohn and Stoltz in \cite{SpSt}.   

In the hyperbolic time scaling, starting from an initial distribution
associated to some smooth macroscopic initial volume-energy profiles
$(v_0 (y),\mathfrak e_0(y))$, we can prove that these initial profiles
evolve following the linear wave equation $(v(t,y),e(t,y))$ which is
solution of 
$$
\partial_t v = 2a  \partial_y v, \quad \partial_t \mathfrak e 
= a \partial_y (v^2). 
$$
As in Section \ref{sec:hyperb-scal-line} we can introduce the
{\textit{mechanical energy}}  $\mathfrak e_{mech} (y,t) =\tfrac{v^2 (y,t)}{2}$ and
the thermal energy  $T(y,t) =  \mathfrak e(y,t) - \mathfrak e_{mech}
(y,t)$. The later remains constant in time.

Mutatis mutandis the discussion of Sections
\ref{sec:superd-evol-temp}, \ref{sec:diff-behav-phon} and
\ref{sec:equil-fluct} can be applied to this model with two conserved
quantities with very similar conclusions {\footnote{In \cite{bgj} only
    the equilibrium fluctuations are considered but the methods
    developed in \cite{jko2}, \cite{koyper} can be applied also to the
    models considered in this section.}} (\cite{bgj, bgjss}). The interesting
difference is that in (\ref{eq:fhe}) and (\ref{eq:11}) the fractional
Laplacian has to be replaced by the {\textit{skew fractional
    Laplacian}}:  
\begin{equation}
\label{eq:fhe2cons}
 \partial_t \frak T = -c\, \{ \, |\Delta_y|^{3/4} -\nabla_y
 |\Delta_y|^{1/4}\} \; \frak T, \qquad \frak T(y,0) 
  =  T(y,0)
\end{equation}
for a suitable explicit constant $c>0$. The skewness is produced here by the interaction of the (unique) sound mode with the heat mode. In the models of Section \ref{sec:introduction} which conserve three quantities, there are two sound modes with opposite velocities. The skewness produced by each of them is 
exactly counterbalanced by the other one so that it is not seen in the final equations (\ref{eq:fhe}) and (\ref{eq:11}).   

\paragraph{The extension problem for the skew--fractional Laplacian}
\label{subsec:extension}
For the model with two conserved quantities introduced in this
section,  the derivation of the skew fractional heat
equation, at least at the level of the fluctuations in equilibrium as
defined in Section \ref{sec:equil-fluct}, can be implemented by means
of the so-called 
{\em extension problem} for the fractional Laplacian
\cite{StroockVaradhan}, \cite{CaffarelliSilvestre}. 
As we will see, this extension problem does not only {provides} a
different derivation, but it also clarifies the role
of the other (fast) conservation law (i.e. the volume). 
It can be checked that for any $\beta >0$ and any
$\rho \in \mathbb R$, the product measure with Gaussian marginals of
mean $\rho$ and variance (temperature) $\beta^{-1}$ are stationary under the
dynamics of $\{\eta_x(t); x \in \mathbb Z\}$. Let us assume that the
dynamics starts from a stationary state. For simplicity we assume
$\rho =0$ and $a=1$. {The space-time energy
correlation function $S_\epsilon(x,t)$  is defined} here by 
\[
S_\epsilon(x,t) = \left\langle (\eta_x(\epsilon^{-3/2} t)^2-\beta^{-1})\, (\eta_0(0)^2-\beta^{-1})\right\rangle.
\]
It turns out that the energy fluctuations are driven by volume
correlations. Therefore it makes sense to define the {\em volume}
correlation function as 
\[
G_\epsilon(x,y,t) = \left\langle \eta_{ x} (\epsilon^{-3/2} t) \eta_{ y}(\epsilon^{-3/2} t)\;  (\eta_0(0)^2 - \beta^{-1})\right\rangle.
\]
Let $f:[0,T] \times \mathbb R \to \mathbb R$ be a smooth, regular
function and for each $t \in [0,T]$, let $u_t: \mathbb R \times
\mathbb R_+ \to \mathbb R$ be the solution of the boundary-value
problem 
\[
\left\{
\begin{array}{r @{\;=\;}l}
-\partial_x u + \gamma \partial_y^2 u & 0\\
\partial_yu(x,0) & \partial_x f(x,t).
\end{array}
\right.
\]
It turns out that $\partial_x u_t(x,0) = \mc L f_t(x)$, where $\mc L$
is the skew fractional Laplacian defined in \eqref{eq:fhe2cons}.  

For test functions $f: \mathbb R \to \mathbb R$ and $u: \mathbb R
\times \mathbb R_+ \to \mathbb  R$ define 
\[
\langle S_\epsilon(t), f\rangle_\epsilon =  \sum_{x \in \mathbb Z} S_\epsilon(x,t) f(\epsilon x),
\]
\[
\langle G_\epsilon(t), u\rangle_\epsilon = \sum_{\substack{ x,y \in \mathbb Z \\ x<y}} G_\epsilon(x,y,t)\;  u\big(\tfrac{\epsilon}{2} (x+y), \sqrt{\epsilon} (y-x)\big).
\]
After an explicit calculation we have:
\[
\tfrac{d}{dt} \Big\{ \langle S_\epsilon(t) f_t\rangle_\epsilon
-\tfrac{ \epsilon^{1/2}}{2\gamma}  \langle G_\epsilon(t), u_t
\rangle_\epsilon + \epsilon \langle S_\epsilon(t),
u_t(\cdot,0)\rangle_\epsilon\Big\} 
		= \langle S_\epsilon(t), \partial_x f_t \rangle_\epsilon
\] 
plus error terms that vanish as $\epsilon \to 0$. From this
observation it is not very difficult to obtain that,
for any smooth function $f$ of compact support,
\[
\lim_{\epsilon \to 0} \langle S_\epsilon(t), f\rangle_\epsilon = \int P(t,x) f(x) dx,
\]
where $P(t,x)$ is the fundamental solution of the skew fractional heat
equation {\eqref{eq:fhe2cons}}. 

Let us explain in more details why the introduction of the test
function $u_t$ solves the equation for the energy correlation function
$S_\epsilon(x,t)$. It is reasonable to parametrize volume correlations
by its distance to the diagonal $x=y$. The microscopic current
associated to the energy $\eta_x(t)^2$ is equal to
$\eta_x(t)\eta_{x+1}(t)$, which can be understood as the volume
correlations around the diagonal $x=y$. Volume evolves in the
hyperbolic scale with speed $2$. Fluctuations around this transport
evolution appear in the diffusive scale and are governed by a
diffusion equation. This means that at the hyperbolic scale
$\epsilon^{-1}$, fluctuations are of order $\epsilon^{-1/2}$,
explaining the non-isotropic space scaling introduced in the
definition of $\langle G_\epsilon(t), u\rangle_\epsilon$. It turns out
that the couple $(S_\epsilon, G_\epsilon)$ satisfies a closed system
of equations, which can be checked to be a semidiscrete approximation
of the system 
\[
\left\{
\begin{array}{r@{\;=\;}l}
\sqrt{\epsilon} \partial_t u_t & -\partial_x u_t + \gamma \partial_y^2 u_t\\
\partial_y u_t(x,0) & \partial_x f_t(x) \\
\partial_t f_t & \partial_x u_t(x,0).
\end{array}
\right.
\]
Therefore, the volume serves as a fast variable for the evolution of the energy, which corresponds to a slow variable. The extension problem plays the role of the cell problem for the homogenization of this fast-slow system of evolutions.

\section{The dynamics in higher dimension}
\label{sec:model-d-dimensions}

One of the interesting features of the harmonic dynamics with energy
and momentum conservative noise is that they reproduce, at least
qualitatively, the expected behavior of the non-linear dynamics,
also in higher dimensions. In particular in the three or higher dimension
the thermal conductivity, computed by the Green-Kubo formula, is finite,
while it diverges logarithmically in two dimension (always for
non-acoustic systems), cf. \cite{bborev, bbo2}.  

In dimension $d\ge 3$ it can be also proven that equilibrium
fluctuations evolve diffusively, i.e. that the asymptotic correlation 
$\tilde S^{33}(y,t)$, defined in section \pageref{sec:equil-fluct} but
with a diffusive scaling, satisfies \cite{bo}:
\begin{equation}
  \label{eq:20}
  \partial_t \tilde S^{33} = D \partial_y^2 \tilde S^{33}
\end{equation}
where $D$ can be computed explicitely by Green-Kubo formula in terms of
$\omega$ and scattering rate $R$ \cite{bborev, bbo2}.
Similar finite diffusivity and diffusive evolution of the fluctuations
are proven for pinned models ($\hat\alpha(0) >0$), see \cite{bo}. 
In two dimensions, while the logarithmic divergence of the Green-Kubo
expression of the thermal conductivity is proven in \cite{bbo2}, the
corresponding diffusive behaviour at the logarithmic corrected time
scale is still an open problem. For the result obtained from the
kinetic equation see  \cite{Ba}. The two-dimensional model is
particularly interesting in light of the large thermal
conductivity measured experimentally on graphene \cite{graphene}, an
essentially a two-dimensional material. 

\section{Thermal boundary conditions and the non-equilibrium
  stationary states}
\label{sec:therm-bound-cond}

Traditionally the problem of thermal conductivity has been approached by
considering the stationary non-equilibrium state for a finite system
in contact with heat bath at different 
temperatures (cf.\cite{rll, sll}). 
This set-up is particularly  suitable for numerical simulations and 
convenient because it gives an straight
operational definition of the thermal conductivity in terms of the
stationary flux of energy, avoiding to specify the macroscopic
evolution equation. But for the theoretical understanding and
corresponding mathematical proofs of the
thermal conductivity phenomena, this is much harder than the
non-stationary approach described in the previous section. This because
the stationary state {conceals} the space-time scale.      

The finite system {consists of}
$2N+1$ atoms, labelled by $x=-N,\dots,N$, with end points connected to
two heat baths 
at temperature $T_l$ and 
$T_r$. These  baths are modeled by  Langevin stochastic dynamics,
so that the evolution equations are given by 
 \begin{equation}
  \label{eq:sdeN-fin}
  \begin{split}
    \dot q_x &= p_x, \qquad \qquad x=-N,\dots,N,\\
    \dot p_x &= -(\alpha_N * q(t))_x + \left(p_{x+1}(t^-) -
      p_x(t^-)\right) \dot N_{x,x+1}(t) + \left(p_{x-1}(t^-) -
      p_x(t^-)\right) \dot N_{x-1,x}(t)\\
    &\quad + \delta_{x,-N} \left( - p_{-N} + \sqrt{2T_l}\; dw_{-N}(t)\right) 
    + \delta_{x,N} \left( - p_N + \sqrt{2T_r}\; dw_N(t)\right) ,
  \end{split}
\end{equation}
where $w_{-N}(t), w_N(t)$ are two independent standard Brownian motions,
and the coupling $\alpha_N$ is properly defined in order to take into account
the boundary conditions
. 
For this finite dynamics there is a
(non-equilibrium) unique stationary state, {where the energy flow from the
hot to the cold side.} Observe that because of the exchange noise
between the atoms, the stationary state is not Gaussian, unlike in the case studied in \cite{rll}. 
  
Denoting the stationary energy flux by $J_N$,
the thermal conductivity of the finite chain is defined as
\begin{equation}
  \label{eq:TCN}
  \kappa_N \ =\  \lim_{|T_l - T_r|\to 0}\ \frac{(2N+1) J_N}{T_l - T_r} .
\end{equation}
 For a finite $N$ it is not hard to prove that $\kappa_N$ can be expressed in
terms of the corresponding Green-Kubo formula.
 For the periodic
boundary unpinned acoustic case this identification gives
$\kappa_N \sim N^{1/2}$ 
(cf. \cite{bborev}).  For the noise that conserves only the energy, but not
momentum (like independent random flips of the signs of the momenta),
the system has a finite thermal diffusivity and the limit $\kappa_N\to
\kappa$, {as $N\to+\infty$,} can be computed explicitely, as proven in
\cite{bo06}. 

The natural question is about the macroscopic evolution of the
temperature profile in a non-stationary situation, and the
corresponding stationary profile. It turns out that this macroscopic
equation is given by a fractional heat equation similar to
\eqref{eq:super1} with a proper definition of the fractional laplacian
$|\Delta|^{3/4}$ on the interval $[-1,1]$ subject to the boundary conditions
$\frak T(-1) = T_l, \frak T(1) = T_r$. 
This is defined by using the following orthonormal basis of functions
on the interval $[-1,1]$: 
\begin{equation}
  \label{eq:ONb}
  u_n(y) = \cos\left( n\pi(y+1)/2 \right) ,\qquad {n=0,1,\ldots}
\end{equation}
Any continuos function $f(y)$ on $[-1,1]$ can be expressed 
{ in terms of a series expansion in 
$u_n$. } 
Then we define $|\Delta|^s u_n (y) = (n\pi/2)^{2s}
u_n(y)$. Observe that for $s=1$ we recover the usual definition of the
Laplacian. {For $s\neq 1$ this is \emph{not} equivalent 
to other definitions of the fractional Laplacian in a bounded
interval, e.g. \cite{zlr}}. 

  Correspondingly the stationary profile is given by 
\begin{equation}\label{ssprof}
\begin{split}
&|\Delta|^s \frak T=0, \;\mbox{in }(-1,1)\\
&\frak T(-1)=T_l,\, \frak T(1)=T_r,
\end{split}
\end{equation}
namely 
$\frak T(y)=\frac 1 2 (T_l+T_r)+\frac 1 2 (T_l-T_r)\theta(y)$, with
$$
\theta(y)=c_s\sum_{m\,\text{odd}}\frac 1 {m^{2s}} \cos\big(m\pi(y+1)/2\big),
$$
$c_s$ such that $\theta(-1)=1$. This expression corresponds with the
{formula} computed directly in \cite{lmp} using a continuous
approximation of the 
covariance matrix of the stationary state for the dynamics with fixed
boundaries.

\section{The non-linear chain}
\label{sec:non-linear-chain}

From the above rigorous results on the harmonic chain with the random
collision dynamics, and the arguments of Spohn from fluctuating
hydrodynamics and mode couplings (cf. \cite{Sp13} and \cite{spohn-here}),
we can conjecture the corresponding behaviour in the anharmonic case.

Let us consider just nearest neighbor interaction given by the
potential energy $V( q_x-  q_{x-1})$ of an anharmonic
spring. We assume $V:\bbR\to(0,+\infty)$ is smooth
  and that it grows quadratically at infinity.
Define the energy of the oscillator $x$ as
\begin{equation}
\label{011705}
 e_x ({ r}, { p}):= \frac{p_x^2}{2}+ V(r_x) 
\end{equation}
The dynamics is defined as the solution of {the Newton} equations
\begin{equation}
\dot { q}_{x}(t)=  p_x, \quad  \dot {p}_x(t)=- \left(V'(r_x) - V'(r_{x+1})\right), 
\quad x\in\bbZ
\label{eq:bash}
\end{equation}
plus a random exchange of velocities as in {the
  previous} sections, 
regulated by an intensity $\gamma$. 
The equilibrium Gibbs measures are parametrized by
$$
\boldsymbol\lambda= (\beta^{-1}\ \text{(temperature)}, \bar p\
\text{(velocity)}, \tau\ \text{(tension)}),
$$
 and are given explicitly by 
\begin{equation}
  \label{eq:gibbs}
  d\nu_{\boldsymbol\lambda} = \prod_x e^{-\beta (V(r_x)  -\tau r_x
    +\tfrac{(p_x-{\bar p})^2}{2}) 
   - \mathcal G(\boldsymbol\lambda)} \; dr_x \; dp_x 
\end{equation}
When a random exchange of velocity is present ($\gamma >0$) it can be
proven that these are the only \emph{regular} translation invariant
stationary measures (\cite{FFL, bo14}).
We have 
\begin{equation*}
\nu_\lambda (p_x) ={\bar p}, \quad \nu_{\lambda} (r_x) =-\frac1\beta \partial_{\tau} {\mathcal G}:={\bar r}, \quad \nu_{\lambda} (e_x) = \frac1\beta -\partial_{\beta} \mathcal G -\frac{\tau}{\beta} \partial_{\tau} {\mathcal G}:={\bar e}.
\end{equation*}
These thermodynamical relations can be inverted to express the
parameters $(\beta^{-1}, {\bar p}, \tau)$ in terms of $(\bar p, \bar
r, \bar e)$. It turns out that the tension is then equal to a
nonlinear function ${\bar{\tau}} ({\bar r}, {\bar u})$ of the average
stretch $\bar r$ and the average internal energy ${\bar u} ={\bar e} -
{\bar p}^2/2$. 

After the hyperbolic rescaling of the dynamics, we expect that the
empirical distribution of the balanced quantities converge to the
system of hyperbolic equations:
\begin{equation}
  \label{eq:euler}
  \begin{split}
    \partial_t \bar r &= \partial_y \bar p\\
    \partial_t \bar p &= \partial_y \bar \tau\\
    \partial_t \bar e &= \partial_y \left(\bar p{\bar \tau}\right).
      \end{split}
\end{equation}

This limit can be proven, under certain condition on the
boundaries, in the smooth regime, if the microscopic dynamics is
perturbed by a random exchange of velocities between nearest neighbors
particles particles by using relative entropy methods (\cite{OVY, EO, bo14}). 
{The limit should be still valid after  shocks develop}, with
the limit profile given by an entropic weak solution. This is a main
open problem  in hydrodynamic limits
.

After a long time, the {(entropy)} solution of \eqref{eq:euler} should
converge (maybe in a weak sense) to some mechanical equilibrium
characterized by: 
\begin{equation}
  \label{eq:mechequi}
  \bar p(y) = p_0, \qquad {\bar \tau}(\bar r(y), \bar u(y)) = \tau_0,\ \ 
  \text{{for some constants $p_0$, $\tau_0$}}. 
\end{equation}
It is very hard to characterize all possible stationary
{solutions} that 
satisfy \eqref{eq:mechequi}. {Probably} they are generically very
irregular. But certainly if we start with {a smooth initial condition 
that satisfies} \eqref{eq:mechequi}, they do not move. Also by
{the relative entropy methods}, it is possible to prove that starting with
such initial profiles, the empirical distribution of the balanced
quantities will converge {at} the hyperbolic {space-time} scale to such
a stationary solution at any time. 


Still we do know that the microscopic dynamics will converge to a global
equilibrium, 
so this implies that there exists a larger time scale such that 
these profiles will evolve and eventually reach also thermal
equilibrium. 

There is a numerical evidence and heuristic arguments about the
divergence of the Green-Kubo formula defining the thermal diffusivity
{for such} one dimensional systems, so we expect that the larger time
scale {at which these profiles evolve} is superdiffusive. 

From the nonlinear fluctuation hydrodynamics (\cite{Sp13}), one can conjecture
the following: the space-time scale is $(\eps^{-1}x,\eps^{-2a}
t)$, and the temperature $T(x,t) = \beta^{-1}(x,t)$ evolves 
following some fractional heat equation, possibly non-linear. 
If $V$ is symmetric and  $\tau_0= 0$, then $a = 3/4$, and in all
other cases $a = 5/6$.

\section{The disordered chain}
\label{sec:disorder}

The effect of disorder on transport and phonons localization
properties in chains of oscillators has attracted a lot of interest
(\cite{dhar,sll}). Randomness can appear at the level of the masses of
particles or at the level of the potentials. We consider only the case
of random masses with non random potential $V$ or the case of non
random masses and non random interaction potential $V$ with random
harmonic pinnings. In the first case, the Hamiltonian is then given by 
$$
{\mc H} ( p,q):= \sum_{x} \cfrac{p_x^2}{2m_x} + \sum_{x, x'}  V (q_x, q_x')
$$ 
where $\{m_x\}$ are positive random variables, while in the second 
case, the Hamiltonian is given by  
 $$
{\mc H} ( p,q):= \sum_{x} \cfrac{p_x^2}{2m} 
+ \sum_{x,x'}  V (q_x ,q_x') + \sum_x \nu_x q_x^2
$$
 where $\{\nu_x\}$ are positive random variables, $m>0$ being the mass
 of the particles. The presence of  randomness is relevant for the
 thermal properties of the system but the fact that randomness affect
 potentials or masses is not.  
 
For one dimensional unpinned disordered harmonic chains it is known
that the behavior of the conductivity is very sensitive to the
boundary conditions since it can diverge as $\sqrt{N}$ or vanish as
$1/\sqrt{N}$ with the systems length $N$ (\cite{AH, CL, RG, V}). If
harmonic pinning is added localization of normal modes leads an
exponential decay of the heat current and a zero conductivity. The
situation in higher dimensions, even in the case of harmonic
interactions, is still under debate but it is expected that
conductivity is finite in dimension $d\ge 3$ if disorder is
sufficiently weak (\cite{KCRDLS}). About the effect of nonlinearities,
numerical evidences suggest that a very small amount of anharmonicity
in pinned chains is sufficient to restore a diffusive regime with a
positive finite value  of the conductivity (\cite{DL}). However it is
a challenging open question to decide if the transition from an
insulator to a conductor occurs at zero or some finite small value of
anharmonicity (\cite{Basko, DL, OPH}).  

It is suggestive to think that a stochastic noise could affect
transport properties of harmonic chains in some rough sense similar to
the addition of nonlinearities. This question has been first address
in \cite{B0} in the Green-Kubo formula setting, revisited in
\cite{DVL} from the non equilibrium stationary state point of view
(see section \ref{sec:therm-bound-cond}) and extended in \cite{BH,H1} to
incorporate weakly nonlinear chains. In these papers, the authors
consider a disordered harmonic chain, or weakly nonlinear in \cite{BH,
  H1}, with a stochastic noise which consists to flip, independently
for each particle, at independent random exponential times of mean
$1/\lambda$, $\lambda>0$, the velocity of the particle. Notice that
this energy conserving noise is very different from the noise
considered in the rest of the paper since it does not conserve
momentum. In particular, for ordered pinned and unpinned nonlinear
chains, this noise is sufficient to provide a finite conductivity
$\kappa$ \cite{BO4}. However, it turns out that for an ordered
harmonic chain, $\kappa \sim \lambda^{-1}$ and, as expected, increases
to infinity as the strength of the noise $\lambda \downarrow 0$.  In
\cite{BH, H1} it is proved that localization effects persist:
$\kappa={\mc O} (\lambda)$ for a pinned disordered chain with a small
anharmonic potential, and $\kappa \sim \lambda$ for a pinned harmonic
chain. As far as we know disordered chains with energy-momentum
conserving noise have never been investigated.



%
\begin{acknowledgement}
  We thank Herbert Spohn for many inspiring discussions on this
  subject. 
 
  The research of C\'edric Bernardin was supported in part by the French
  Ministry of 
  Education through the grant ANR-EDNHS.  
  The work of Stefano Olla has been partially supported by the
  European Advanced Grant {\em Macroscopic Laws and Dynamical Systems}
  (MALADY) (ERC AdG 246953) and by a CNPq grant \emph{Sciences Without
  Frontiers}. Tomasz Komorowski acknowledges the support of the
  Polish National Science Center grant UMO-2012/07/B/SR1/03320. 
\end{acknowledgement}
%



\begin{thebibliography}{99.}%
%
%

\bibitem{AH} O. Ajanki, F. Huveneers, Rigorous scaling law for the heat current in disordered harmonic chain Commun. Math. Phys. {\bf 301} 841Ð83 (2011).


\bibitem{bborev} G. Basile, C. Bernardin, S. Olla, {A momentum
  conserving model with anomalous thermal conductivity in low
  dimension}, Phys. Rev. Lett. \textbf{96}, 204303 (2006),
DOI 10.1103/PhysRevLett.96.204303.

\bibitem{bbo2}  G. Basile, C. Bernardin, S. Olla, {Thermal Conductivity
  for a Momentum Conservative Model}, Comm. Math. Phys. \textbf{287},
  67--98, (2009).

\bibitem{babo} G. Basile, A. Bovier, {Convergence of a kinetic
    equation to a fractional diffusion equation,} 
 Markov Proc. Rel. Fields \textbf{16}, 15-44 (2010);


\bibitem{bo} G. Basile, S. Olla, {Energy Diffusion in Harmonic System
  with Conservative Noise}, J. Stat. Phys., \textbf{155}, no. 6,
  1126-1142, (2014), (DOI) 10.1007/s10955-013-0908-4.

\bibitem{bkoss} G. Basile, T. Komorowski, S. Olla, Temperature profile
  evolution in a one dimensional chain with boundary thermal
  conditions, in preparation.

\bibitem{BOS} G. Basile, S. Olla, H. Spohn, {Energy transport in
    stochastically perturbed lattice dynamics}, Arch.Rat.Mech.,
   \textbf{195}, no. 1, 171-203, (2009). 
 
\bibitem{Ba}  G. Basile, From a kinetic equation to a diffusion under
an anomalous scaling. Ann. Inst. H. Poincare Prob. Stat., \textbf{50},
No. 4, 1301--1322 (2014), DOI: 10.1214/13-AIHP554. 

\bibitem{Basko} D.M. Basko, Weak chaos in the disordered nonlinear
  Schr\"odinger chain: destruction of Anderson localization by Arnold
  diffusion, Ann. Phys. {\bf{326}} 1577Ð655, (2011). 

\bibitem{B0} C. Bernardin, Thermal conductivity for a noisy disordered
  harmonic chain, J. Stat. Phys. {\bf{133}}(3), 417Ð433 (2008). 


\bibitem{bgj} C. Bernardin, P. Goncalves, M. Jara, {3/4
  Fractional superdiffusion of energy in a system of harmonic
  oscillators perturbed by a conservative noise},
  arxiv.org/abs/1402.1562v3, to appear in
  Arch. Rational Mech. Anal. 2015.    

\bibitem{bgjss} C. Bernardin, P. Goncalves, M. Jara, M. Sasada, M. Simon. From normal diffusion to superdiffusion of energy in the evanescent flip noise limit. J. Stat. Phys. {\bf 159}, no. 6, 1327Ð1368 (2015).

\bibitem{BH} C. Bernardin, F. Huveneers, Small perturbation of a disordered harmonic chain by a noise and an anharmonic potential, Probab. Theory Related Fields {\bf{157}}, no. 1-2, 301Ð331 (2013). 

\bibitem{bo06}  C. Bernardin, S. Olla, Fourier law and fluctuations
  for a microscopic model of heat conduction, J. Stat.Phys., 
  \textbf{118}, nos.3/4, 271-289, (2005). 

 \bibitem{bo14} C. Bernardin, S. Olla, {Thermodynamics and
    non-equilibrium macroscopic dynamics of chains of anharmonic
     oscillators}, Lecture Notes  available at 
   {\tt https://www.ceremade.dauphine.fr/~olla/} (2014), 

\bibitem{BO4} C. Bernardin, S. Olla, Transport Properties of a Chain
  of Anharmonic Oscillators with random flip of velocities, Journal of
  Statistical Physics {\bf{145}}, 1224-1255, (2011).  


\bibitem{BS} C. Bernardin, G. Stoltz, Anomalous diffusion for a class
  of systems with two conserved quantities, Nonlinearity {\bf 25},
  N. 4, 1099-1133 (2012). 

\bibitem{CaffarelliSilvestre} L. Caffarelli, L. Silvestre: An
  extension problem related to the fractional
  Laplacian. Commun. Partial Differ. Equations \textbf{32}(8),
  1245-1260 (2007). 

\bibitem{CL} A. Casher, J.L. Lebowitz, Heat flow in regular and disordered harmonic chains, J.Math.Phys.{\bf{12}} 1701Ð11 (1971).

\bibitem{dhar} A. Dhar, Heat Transport in Low Dimensional Systems,
  Adv.In Phys. \textbf{57}, 5, 457-537, (2008).

\bibitem{DL} A. Dhar and J.L. Lebowitz, Effect of phonon-phonon interactions on localization Phys.Rev.Lett.{\bf{100}} 134301 (2008).

\bibitem{DVL} A. Dhar, K. Venkateshan, J.L. Lebowitz,  Heat conduction in disordered harmonic lattices with energy- conserving noise. Phys. Rev. E {\bf{83}}(2), 021108 (2011).

\bibitem{EO} N. Even, S. Olla, {Hydrodynamic Limit for an Hamiltonian
  System with Boundary Conditions and Conservative Noise},
  Arch.Rat.Mech.Appl., \textbf{213}, 561-585,  (2014). DOI
  10.1007/s00205-014-0741-1 

\bibitem{FFL} J. Fritz, T. Funaki, J.L. Lebowitz, Stationary states of
  random Hamiltonian systems, Probab. Theory Related Fields, {\bf 99}
 , 211--236 (1994). 

\bibitem{H1} F. Huveneers, Drastic fall-off of the thermal conductivity for disordered lattices in the limit of weak anharmonic interactions, Nonlinearity {\bf 26}, no. 3, 837Ð854 (2013). 

\bibitem{jko} M. Jara, T. Komorowski, S. Olla, {A limit theorem for an
  additive functionals of Markov chains}, Annals of Applied
Probability \textbf{19}, No. 6, 2270-2300, (2009).

\bibitem{jko2}  M. Jara, T. Komorowski, S. Olla, {Superdiffusion of
    Energy in a Chain of harmonic Oscillators with
    Noise}, Commun. Math. Phys. \textbf{339}, 407--453 (2015), (DOI)
  10.1007/s00220-015-2417-6. 

\bibitem{koyper}  T. Komorowski, S. Olla, {Ballistic and
    superdiffusive scales in macroscopic evolution of a chain of
    oscillators}, arXiv:1506.06465, (2015). 

\bibitem{kononac}  T. Komorowski, S. Olla, {Diffusive propagation of
    energy in a non-acoustic chain}, in preparation.



\bibitem{KOR} T. Komorowski, S. Olla, L. Ryzhik, {Asymptotics of the
    solutions of the  stochastic  lattice   wave equation},
  Arch. Rational Mech. Anal.,  {\bf 209}, 455-494, 2013. 


\bibitem{KS} T. Komorowski, L. Stepien, {Long time, large scale
    limit of the Wigner transform for a system of linear oscillators
    in one dimension}, Journ. Stat. Phys., Vol. {\bf  148},   pp 1-37
  (2012). 
  
\bibitem{KCRDLS} A. Kundu, A. Chaudhuri, D. Roy, A. Dhar,
  J.L. Lebowitz, H. Spohn, {Heat transport and phonon localization in
    mass-disordered harmonic crystals}, Phys. Rev. B {\bf 81} 064301,
  2010   
 
\bibitem{lll} O. Lanford, J.L. Lebowitz, E. Lieb, {Time Evolution of
  Infinite Anharmonic Systems}, J. Stat. Phys. \textbf{16}, n. 6,
  453-461, (1977).

\bibitem{sll} S. Lepri, R. Livi, A. Politi,  {Thermal Conduction in
 classical low-dimensional lattices}, Phys. Rep. \textbf{377}, 1-80 (2003).

\bibitem{llp97} S. Lepri, R. Livi, A. Politi, {Heat conduction in
    chains of nonlinear oscillators}, 
  Phys. Rev. Lett. \textbf{78}, 1896 (1997).

\bibitem{lmp} S. Lepri, C. Meija-Monasterio, A. Politi, A stochastic
  model of anomalous heat transport: analytical solution of  the
  steady state, J. Phys. A: Math. Gen. \textbf{42}, (2009) 025001.

\bibitem{LS} Lukkarinen, J.,  Spohn, H. (2006). {Kinetic Limit for
    Wave Propagation in a Random Medium.} Arch. for Rat.
  Mech. and Anal., \textbf{183}, 1,
  93-162. 

\bibitem{Luk-spohn} Lukkarinen, J. and Spohn, H., Anomalous
  energy transport in the FPU-$\beta$ chain. Comm. Pure Appl. Math., 61:
  1753--1786, (2008). doi: 10.1002/cpa.20243

\bibitem{mmm} A. Mellet, S. Mischler, C. Mouhot, {Fractional
    diffusion limit for collisional kinetic equations},  Arch. for
  Rat. Mech. and Anal. \textbf{199},  2, pp 493-525 (2011)

\bibitem{OPH} V. Oganesyan, A. Pal, D. Huse, Energy transport in
  disordered classical spin chains, Phys.Rev.B {\bf{80}}, 115104
  (2009). 

\bibitem{OVY} S. Olla, S.R.S.~Varadhan, H.~T.~Yau, 
{Hydrodynamic Limit for a Hamiltonian System 
with Weak Noise}, {Comm. Math. Phys. \textbf{155}}, 523-560, 1993.

\bibitem{Pe} R. E. Peierls, 
{Zur kinetischen Theorie der Waermeleitung in Kristallen},
{Ann. Phys. Lpz. \textbf{3}},1055-1101, (1929).


 \bibitem{rll} Z. Rieder, J.L. Lebowitz, E. Lieb,  { Properties
     of   harmonic crystal in a stationary non-equilibrium state},
 J. Math. Phys. \textbf{8}, 1073-1078 (1967).
 
 \bibitem{Sp06} H. Spohn {The phonon Boltzmann equation, properties
     and link to weakly anharmonic lattice dynamics}, 
{J. Stat. Phys. \textbf{124}}, no. 2-4, 1041-1104, (2006).

\bibitem{Sp13} H. Spohn, {Nonlinear fluctuating hydrodynamics for
  anharmonic chains},  J. Stat. Phys. \textbf{154} no. 5, 1191-1227, (2014).

\bibitem{spohn-here}  H. Spohn, Fluctuating hydrodynamics approach to
equilibrium time correlations for anharmonic chains, this issue.

\bibitem{SpSt} H. Spohn, G. Stoltz, Nonlinear fluctuating
  hydrodynamics in one dimension: the case of two conserved fields,
  J. Stat. Phys., \textbf{160}, 861--884 (2015), DOI 10.1007/s10955-015-1214-0. 

\bibitem{StroockVaradhan} D. W. Stroock, S.R.S. Varadhan, Diffusion
  Processes with Boundary Conditions, Comm. Pure Appl. Math. \textbf{24},
  147-225 (1971). 

\bibitem{RG} R.J. Rubin, W.L. Greer, Abnormal lattice thermal
  conductivity of a one-dimensional, harmonic, 
isotopically disordered crystal, J. Math. Phys. {\bf 12} 1686Ð701 (1971).


\bibitem{V} T. Verheggen, Transmission coefficient and heat conduction
  of a harmonic chain with random masses: 
asymptotic estimates on products of random matrices, 
Commun. Math. Phys. {\bf{68}}, 69--82 (1979). 

\bibitem{graphene} Y.Xu, Z. Li, W. Duan, Thermal and Thermoelectric
  Properties of Graphene, Small 2014, \textbf{10}, No. 11, 21822199, DOI:
  10.1002/smll.201303701. 

\bibitem{zlr} A. Zola, A. Rosso, M. Kardar, Fractional Laplacian in a
  Bounded Interval, Phys. Rev. E \textbf{76}, 21116 (2007).
\end{thebibliography}
\end{document}